\documentclass[prd, aps, nofootinbib, preprintnumbers, showpacs,
superscriptaddress,twocolumn]{revtex4}

\usepackage{graphicx,epsfig,color}

\usepackage{amssymb,amsmath}
\usepackage[english]{babel}
\usepackage{graphicx}
\usepackage[latin9]{inputenc}
\usepackage{epstopdf}
\usepackage{color}

\newcommand{\jb}{juggling ball}
\newcommand{\jbs}{juggling balls}
\newcommand{\CC}{Caltech-Cornell}

\newcommand{\beq}{\begin{equation}}
\newcommand{\eeq}{\end{equation}}
\newcommand{\beqa}{\begin{eqnarray}}
\newcommand{\eeqa}{\end{eqnarray}}
\newcommand{\ot}{\overline{t}}
\newcommand{\ox}{\overline{x}}
\newcommand{\oy}{\overline{y}}
\newcommand{\oz}{\overline{z}}

\begin{document}

\title{Orbiting binary black hole evolutions with a multipatch high order finite-difference approach}

\author{Enrique~Pazos}
\affiliation{Center for Fundamental Physics, Department of Physics, University of Maryland, College Park, MD
  20742, USA}
  \affiliation{Center for Scientific Computation and Mathematical Modeling, University of Maryland, College Park, MD
  20742, USA}
 \affiliation{Departamento de Matem\'atica, Universidad de San Carlos de Guatemala, Edificio T4, Facultad de Ingenier\'{\i}a, Ciudad Universitaria Z. 12, Guatemala}

\author{Manuel~Tiglio}
\affiliation{Center for Fundamental Physics, Department of Physics, University of Maryland, College Park, MD
  20742, USA}
\affiliation{Center for Scientific Computation and Mathematical Modeling, University of Maryland, College Park, MD
  20742, USA}
  
\author{Matthew D. Duez}
\affiliation{Center for Radiophysics and Space Research, Cornell University, Ithaca, New York, 14853}

\author{Lawrence E. Kidder}
\affiliation{Center for Radiophysics and Space Research, Cornell University, Ithaca, New York, 14853}

\author{Saul A. Teukolsky}
\affiliation{Center for Radiophysics and Space Research, Cornell University, Ithaca, New York, 14853}

\begin{abstract}
We present numerical simulations of orbiting black holes for around twelve cycles,
using a high-order multipatch approach.  Unlike some other approaches,
the computational speed scales almost perfectly for thousands of processors.
Multipatch methods are an alternative to AMR (adaptive mesh refinement), with
benefits of simplicity and better scaling for improving the  resolution in the wave zone.
The results presented here pave the way for multipatch evolutions of black hole-neutron star
and neutron star-neutron star binaries, where high resolution grids are needed to resolve
details of the matter flow.

\end{abstract}

\pacs{04.25.dk, 04.40.Dg, 04.30.Db, 95.30.Sf}

\maketitle

\section{Introduction}

Mergers of binary compact objects (neutron stars or black holes) are
expected to be the main sources of gravitational waves for the ground-based
interferometric detectors LIGO, GEO, VIRGO, and TAMA.  Neutron star-neutron
star and black hole-neutron star binaries are also
interesting because they are leading candidates for explaining the production
of short-duration gamma-ray bursts and because gravitational wave signals
from these events may encode information about the neutron star equation
of state~\cite{Oechslin:2007gn,Read:2009yp,Vallisneri:1999nq}.  Such a
merger can be accurately modeled only by the numerical evolution of the full
Einstein field equations coupled (if a neutron star is present) to an
evolution of the neutron star matter.

Because of advances in numerical relativity in recent years, stable evolutions
can now be performed for most binary cases.  Accuracy and speed are now the
pressing numerical challenges:  how to achieve the minimum error given
limited time and computational resources.  A good code should
converge rapidly with increasing resolution to the exact solution.  Its speed
should scale well with the number of processors used in order to make good use
of parallelization.  Also, an efficient use of resources will require a grid
well adapted to the problem at hand.  This includes using a grid with the most
appropriate shape.  For example, it is reasonable to suppose that excision
inner boundaries and outer boundaries should be spherical.  A good grid will
also use higher resolution where it is most needed.  For example, although the
grid must extend out into the wave zone to extract the gravitational wave
signal, lower resolution is needed in the wave zone than is needed in the
vicinity of a black hole or neutron star.  The need for high resolution in
neutron stars and black hole accretion disks can become particularly acute in
cases of hydrodynamic or magnetohydrodynamic instabilities, such as convective,
Kelvin-Helmholtz, or magnetorotational instabilities.
In such cases, the length scale of
the unstable modes can be much smaller than the radius of the star or disk,
and the evolution will be qualitatively wrong if the instability is completely
unresolved.  

One technique that has been successfully used to deal with this problem
is adaptive mesh refinement (AMR)~\cite{Yamamoto:2008js,Baiotti:2008ra}. 
These AMR codes generally use overlapping Cartesian meshes of varying levels
of refinement, with the finer meshes being used only where they are determined
(by some algorithm) to be needed.  In this paper, we present a different
method of achieving efficient grid coverage, one that is algorithmically
simpler and that possesses some unique advantages.

This different technique for evolving binary compact systems involves using
multiple grid patches, each patch having its own shape, curvilinear coordinates and 
resolution.  The basic ideas behind these multipatch methods have been
worked out in earlier papers~\cite{Lehner:2005bz,Diener:2005tn,Schnetter:2006pg}. 
In these references some particular 
patch configurations using cubes and cubed-spheres were used.  The cubed-sphere
patches were used to construct grids with exactly spherical inner excision
boundaries and outer boundaries.  These methods are, hence, ideal for
calculations that involve excision.  (Using AMR with excision
introduces a number of complications.)  These techniques were then successfully
used to simulate perturbed Kerr black holes~\cite{Dorband:2006gg,Pazos06}.  
Multiple patches in cubed-sphere arrangement have also been used to evolve the
shallow water equations~\cite{ronchi} and to simulate hydrodynamic flows in black
hole accretion disks~\cite{Koldoba:2002kx,Zink:2007xn,Fragile:2008ca}.

Another multipatch approach has been used by the Cornell-Caltech group
to evolve Einstein's equations for binary black hole~\cite{Scheel:2008rj}
and black hole-neutron star~\cite{Duez:2008rb} systems.  In the binary black hole case derivatives
in these simulations are computed pseudospectrally, rather than using finite
differencing.  While pseudospectral methods produce accurate results very
efficiently for binary black hole evolutions, they are much less cost
effective for systems involving matter.  One reason for this is that the
discontinuities that naturally appear in the fluid flow at shocks and
stellar surfaces destroy the exponential convergence of spectral methods. 
In fact, the Cornell-Caltech group found it necessary to evolve the fluid
variables using finite differencing, while evolving the field variables
pseudospectrally.  This required two independent grids: the finite difference
gridpoints used to evolve the fluid, and the collocation points of the pseudospectral
code used to evolve the metric. For the two grids to communicate, variables had to be
interpolated from one grid to the other each timestep, a process which consumed about
one third of the CPU time in each simulation.  
Another problem with pseudospectral techniques is that
they usually do not scale well to large numbers of processors.  In regions
without discontinuities, where spectral convergence is not lost, one cannot,
for example, split one large domain into two domains with half the number of
collocation points each without a significant loss in accuracy.  On the other
hand, accurate simulations of binary neutron star or black hole-neutron star
mergers are not practical without many processors.

It would therefore seem preferable to evolve both the fluid and the metric with finite differencing. 
This could significantly improve the scalability, allowing simulations on hundreds or thousands of
processors.  It would also remove the need for two separate grids and the expensive interpolation between
them.  Multipatch techniques are the natural finite difference version of the Cornell-Caltech pseudospectral
evolution algorithm.  As a first step in that direction, in this paper we evolve a binary black hole system
using multipatches together with high order finite-differencing operators. 
We show that our code converges rapidly, scales well to thousands of processors, and can
stably simulate several orbits of the inspiral.

\section{Evolution equations}
At the continuum level, the techniques used in this paper are exactly those ones previously
used by the \CC\ collaboration in binary black hole simulations. We use the first order form
of the generalized harmonic system presented in~\cite{Lindblom:2005qh}. The evolution variables
in this formulation are the 4-metric $g_{ab}$ and its first derivatives in space and time
$\partial_c g_{ab}$.  (The indices run from 0 to 3.)  We use the constraint preserving boundary
conditions of \cite{Lindblom:2005qh,Rinne:2006vv,Rinne:2007ui}.  The evolution of the gauge is determined
by the gauge source functions $H_{a} = -g^{cd}\Gamma_{acd}$, which are freely specifiable
functions of space and time.  In this paper, the gauge is set by choosing $H_a$ to be constant
in time in a coordinate system that comoves with the holes.  This comoving coordinate system is
determined using the same dual frame and control tracking mechanism as was used for the spectral
binary black hole evolutions~\cite{Scheel:2006gg}.  This technique uses two coordinate frames,
which we label $x^{\bar{\imath}}$ and $x^i$.  The coordinate frame $x^{\bar{\imath}}$
is set to be an asymptotically flat, inertial frame.  All tensor components
are evaluated with respect to this frame.  The gridpoints are fixed in the computational
frame $x^i$.  By means of a mapping between the frames, the computational coordinates
can be made to approximately comove with the system.  For the runs in this paper, we track
the binary using a simple combination of rotation and radial scaling:
\begin{eqnarray}
\label{RotAndScale}
\begin{split}
\ot &= t \\
\ox &= a[x\cos(\theta) - y\sin(\theta)] \\
\oy &= a[x\sin(\theta) + y\cos(\theta)] \\
\oz &= az   \ ,
\end{split}
\end{eqnarray}
where $\theta$ and $a$ are functions of time which are evolved using
a feedback mechanism to keep the location of the black holes
fixed in the computational domain.

The differences between the simulations presented in this paper and the earlier
spectral simulations the type of domain decomposition, and in the numerical techniques
used to compute the right-hand sides of the evolution equations (e.g. how spatial
derivatives are approximated).  Our handling of these issues is described below.

\section{Initial Data}
The initial data that we use here consists of a snapshot at a given time of the 
highest resolution 16-orbit
simulation done by the \CC\ collaboration, which corresponds to the
run 30c1 reported in Refs.~\cite{Boyle:2007ft,Scheel:2008rj}. 
The starting time $t=0$ in our simulations 
corresponds to the instant $t=2887\,M$ of the 16-orbit simulation (with $M$ being the sum of the irreducible masses of 
each black holes). From that point,
the black holes orbit for about 6 orbits before merger,
although our runs stop before the merger takes place.

This way of specifying the initial data has the advantage that there
is no junk radiation present in the computational domain
at our starting time. Since the domains and points used in this paper 
are different from those used in the spectral simulation, we spectrally interpolate 
the initial data to the multipatch domain.

The outer boundary of our domain is a sphere of radius $r=144\, M$. This
value is actually mapped to $r'=105\, M$ by the dual-frame coordinate
transformation, which scales and rotates the inertial coordinates into
the comoving ones. The coordinate transformation is a simple rescaling
of the radial coordinate $r'=a(t)r$ by a time dependent factor, and
a rigid rotation about the $z$ axis. Since the binary system has
been evolving before our $t=0$ time, the scale factor
has a value $a=0.727$ and the rotation angle is $\theta=57.95$ radians at the beginning of our simulations.
The black hole coordinate separation at the beginning of the
30c1 run is $14.44\,M$. At our time $t=0$ the initial coordinate separation 
is $10.5\,M$.

\section{Multi-block domain}
\subsection{Structure}
\begin{figure}
\includegraphics[width=0.45\textwidth]{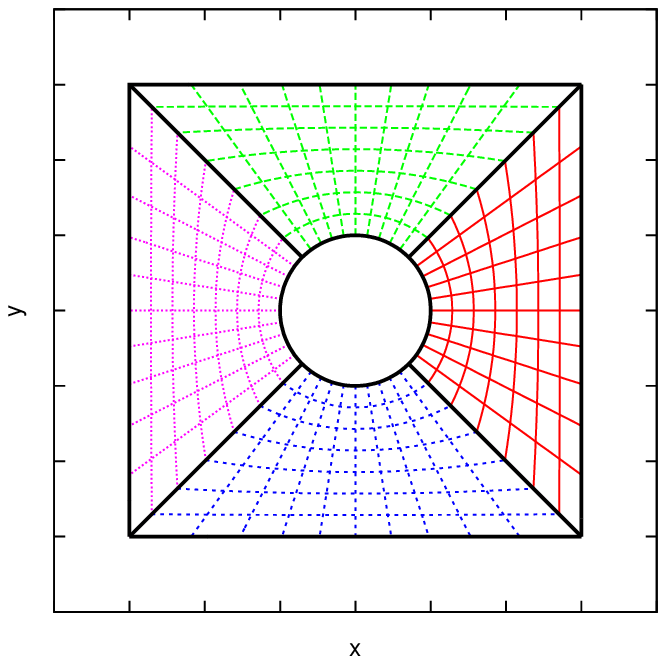}
\includegraphics[width=0.45\textwidth]{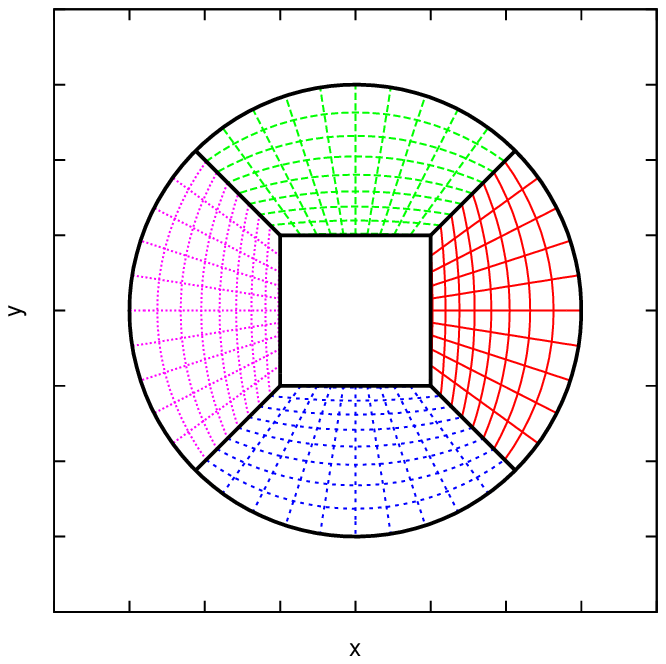}
\caption{Equatorial cross-section of an inner \jb\ (top). Black lines
denote the block boundaries. Colored lines represent the
coordinate grid of each block. Equatorial cross-section of an
outer \jb\ (bottom).}
\label{fig:jugg-in}
\end{figure}

We use two types of basic building patches to cover the whole computational domain.
One is simply a cuboid with a Cartesian coordinate map. The other is
a combination of six patches that we call a {\it \jb}. 
A \jb\ can assume two different configurations. The first of them is  
shown at the top of Fig.~\ref{fig:jugg-in}. It consists of a cube whose interior has
been excised by a sphere. We will refer to it as an {\it inner}
\jb\, because it is the one that we use to excise the interior of each 
black hole and to cover its immediate surroundings. The second
configuration is shown at the bottom of Fig.~\ref{fig:jugg-in} and consists
of a sphere whose interior has been excised by a cube. We will call
it an {\it outer} \jb\, because it is the one that covers the region
away from the black holes, reaching to the outer boundary. Both types
of \jbs\ use a radial coordinate that adjusts smoothly to their
geometry. Each surface of constant radial coordinate is endowed with
six two-dimensional coordinate maps, in the same fashion as the cubed
sphere~\cite{ronchi}.
 In essence, the \jb\ is a collection of six patches,
each of them topologically equivalent to a cube.\footnote{The name
\jb\ was chosen because some real juggling balls have a set of
six quadrilateral-shaped designs on their surface.}

\begin{figure}
\includegraphics[width=0.45\textwidth]{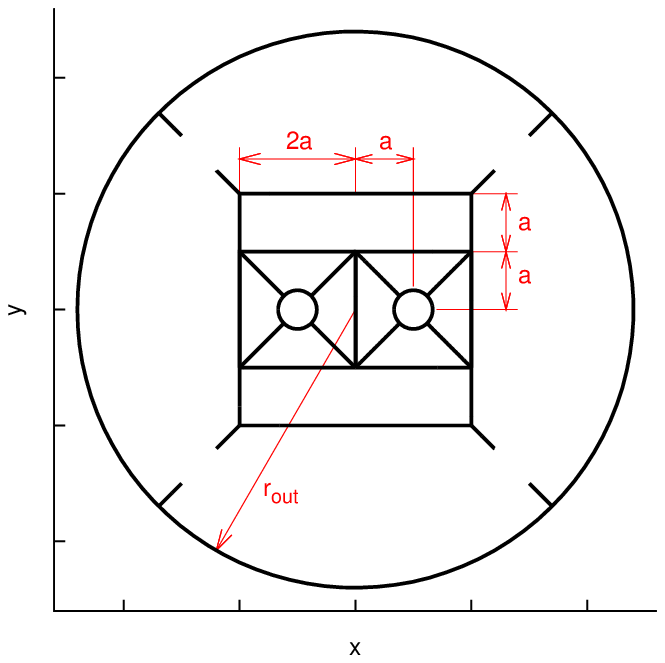}
\includegraphics[width=0.45\textwidth]{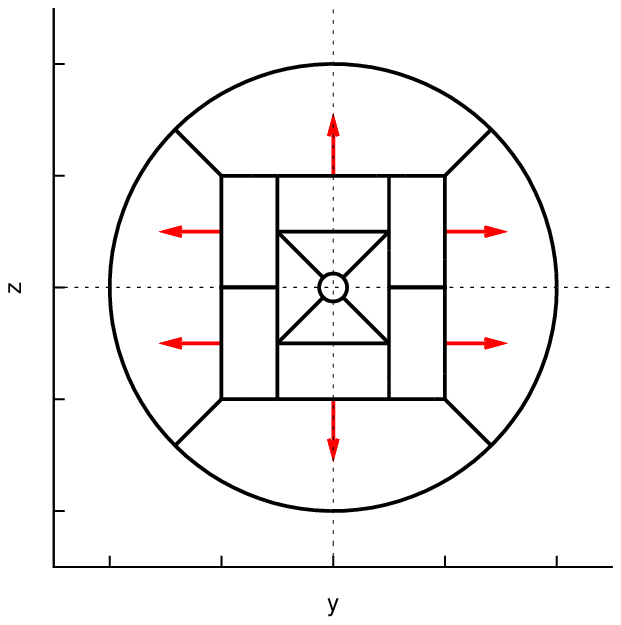}
\caption{Equatorial cut of the computational domain (top).
Schematic figure showing the direction considered as
radial (red arrows) for the cuboidal blocks (bottom).}
\label{fig:domain}
\end{figure}

The basic layout of the full domain used in this paper is shown in Fig.~\ref{fig:domain}.
The centers of the excised spheres (which will be inside each black hole) are located along the $x$ axis at $x=\pm a$.
Here we have used two inner \jbs, one around each black hole. Their 
individual outer boundaries are cubes with sides of length $2a$. 
When they are put together, we end up with a cuboid of dimensions 
$4a \times 2a \times 2a$, with the longest side along the $x$
axis. We surround this structure with six cuboid patches of dimensions
$4a \times 2a \times a$, aligning them along the $y$ and $z$ axes.
After doing so, we end up with a cubical domain with sides of
length $4a$. To complete the patch system we add an outer
\jb\ whose cubical interior holds the two inner \jbs\ plus
the six cuboids. The outer \jb\ enables us to shape the outer
boundary into a sphere, in which case moving the boundary further out requires
an increase in the number of grid points that scales as $O(N)$
(as opposed to $O(N^3)$). 

The total number of patches in this basic configuration is 
6 cuboidal patches  $+$ 6 $\times $ (3 \jbs) $=$ 24 patches.

None of the patches used in this paper overlap with any other (in which case they are usually called {\it blocks}). 
A given block communicates with adjacent ones only by the two-dimensional common
surface between them. Accordingly, we handle parallelization by assigning one block per processor, 
in this way minimizing communication between processors. 

In this basic 24-block
domain case, we would use exactly 24 processors, which is a fairly small number for a binary black 
hole simulation. In order to
achieve higher resolutions by increasing the amount of points per
block, we subdivide the existing blocks into smaller
pieces. Since the topology of each block is cubical, it is straightforward to subdivide them. 
 The guiding principle that we use to accomplish the subdivision is
to keep the same number of points per block for every single
block. Although this condition is not necessary, it is convenient
because it balances the computational load across all the processors.

For the runs presented here, we used 192- and 384-block domains.
The first case is obtained by subdividing the inner \jbs\ uniformly 
in the radial direction 7 times. The 6 cuboids are split by a factor
of 2 and the outer \jb\ is divided 4 times in the radial
and twice in each transverse direction. The 384-block case
is derived from the 192-block one by further split of each block
in the radial direction by a factor of 2.

\begin{figure}
\includegraphics[width=0.45\textwidth]{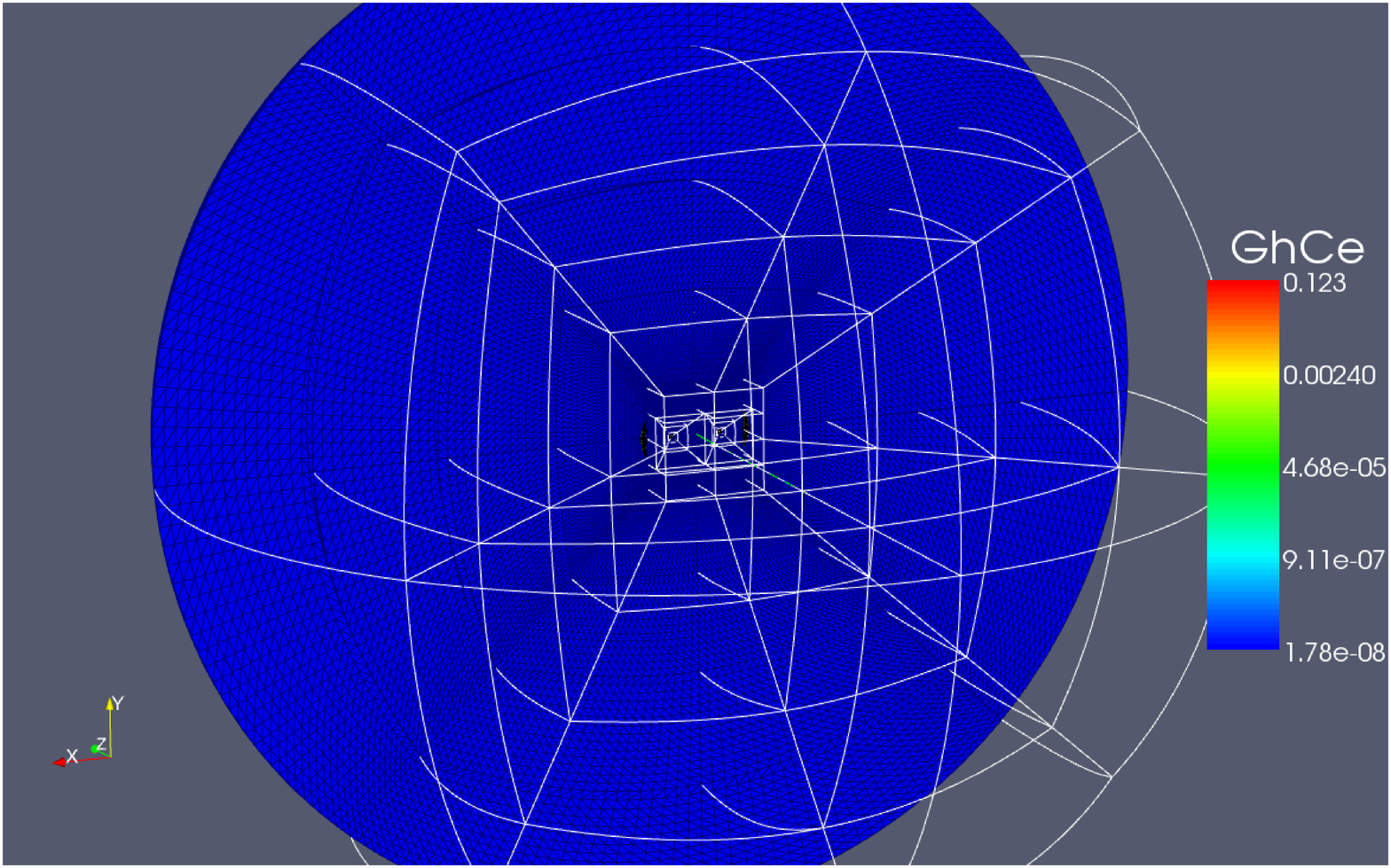}
\includegraphics[width=0.45\textwidth]{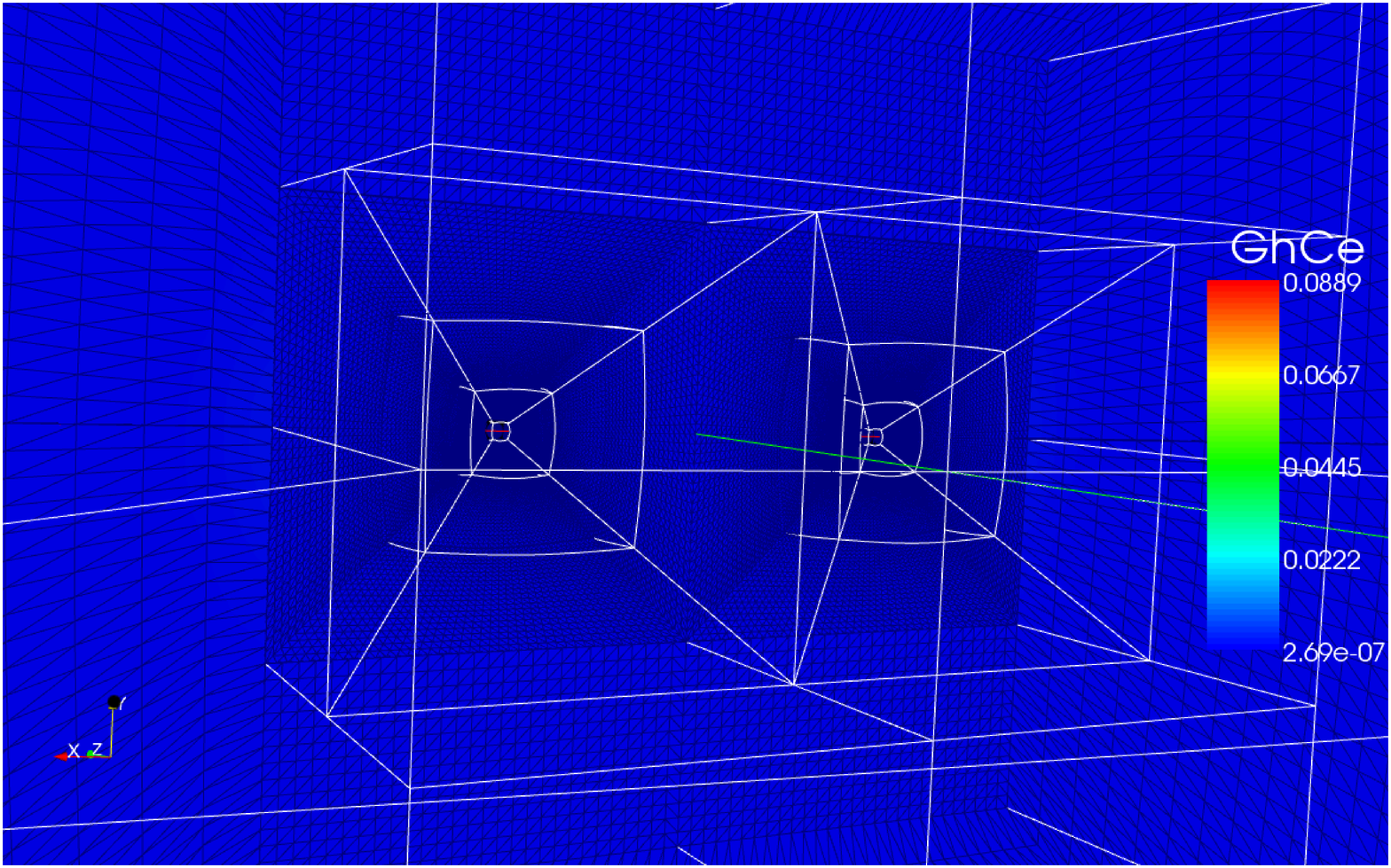}
\includegraphics[width=0.45\textwidth]{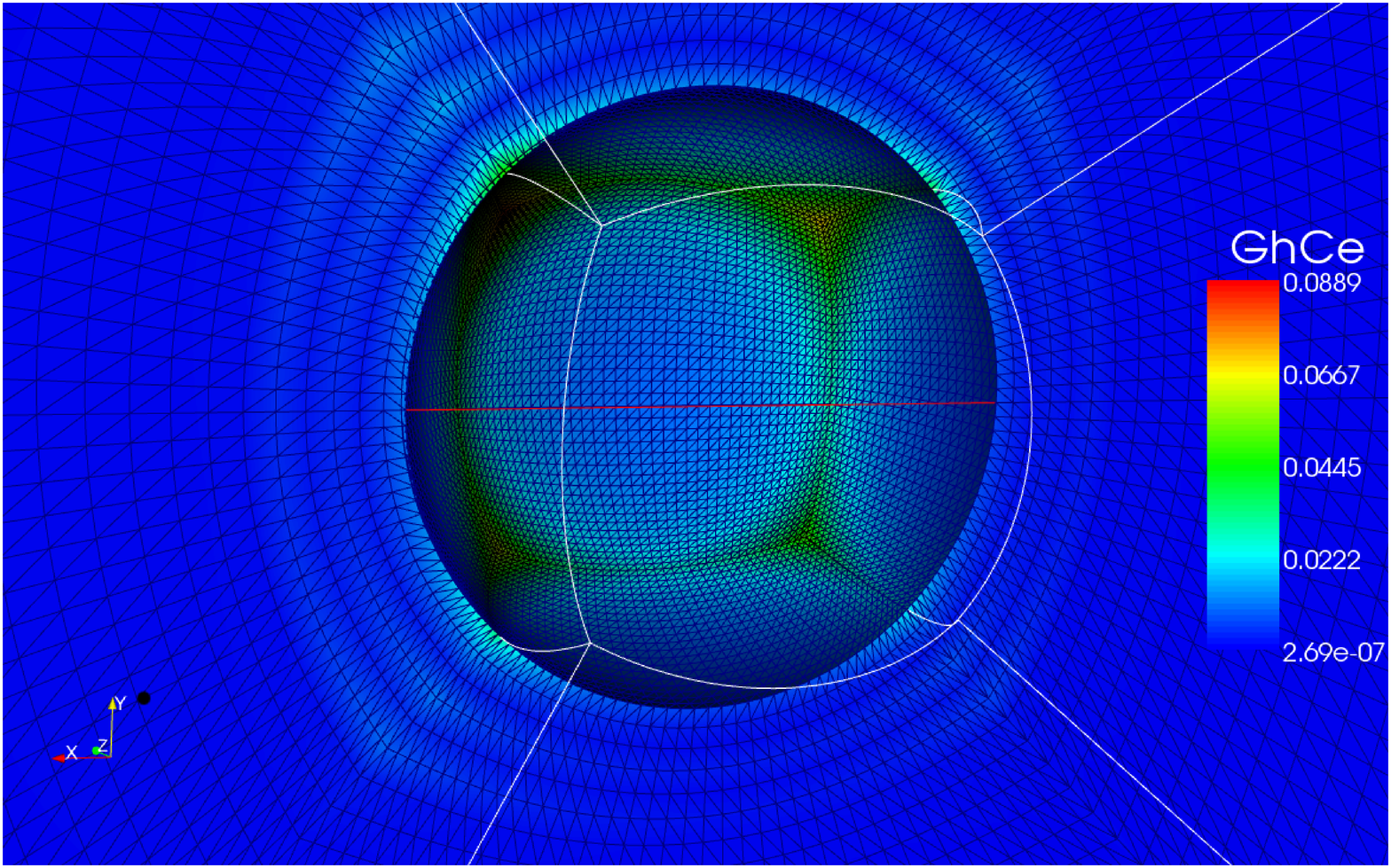}
\caption{Computational domain used in the simulations of this paper}
\label{fig:domain2}
\end{figure}

Figure \ref{fig:domain2} shows the multipatch structures used in this paper. 

\subsection{Numerical techniques} \label{sec:num}

In our simulations we use the $D_{8-4}$ summation-by-parts (SBP) finite difference
operator and its associated dissipation constructed in 
~\cite{Diener:2005tn}. The naming convention is meant to indicate that the derivative is
8th order accurate in the bulk of each block but only 4th order accurate near inter-block boundaries. 
The derivative in the interior of each block is a centered one 
and is modified near boundaries so as to satisfy the SBP property with respect to a diagonal norm; this is 
the cause of the drop in convergence. Information across 
sub-domains is communicated using characteristic variables and a penalty method 
(see \cite{Lehner:2005bz,Diener:2005tn,Schnetter:2006pg} for more details). 

The combination of these techniques guarantees numerical stability, but at the expense of the drop in convergence order 
 near boundaries. For example, in the $D_{8-4}$ case there are eight points near {\em each} boundary 
where the scheme is fourth order. For technical reasons explained below, 
in the simulations of this paper we use a rather large number of blocks and processors (192 and 384),
 with a very small load on each. As a consequence,  the scheme is fourth order nearly everywhere and 
 we expect our simulations to be 4th order convergent. This is indeed what our simulations below show. 
 
\subsection{Resolution}

One of the features that a multipatch method offers is the flexibility
to increase only the radial resolution
while keeping the angular resolution constant. Given that the angular
profile of the waveforms is dominated by a few low-$\ell$ modes,
once a sufficient angular resolution is used the truncation error
will be dominated by the radial resolution.

The approximate 
spherical symmetry in the vicinity of each black hole and
at large distances from them allows the radial direction to be naturally
defined for each \jb\ block. However, for the cuboidal blocks there
is some arbitrariness in how to choose the radial direction. In practice, a
radial direction for these blocks is useful only to define the direction along 
which resolution will be increased.
In Fig.~\ref{fig:domain} the radial directions for the cuboidal
blocks are indicated with arrows.

We use an angular resolution of $\pi/58$ around each black hole and
twice as much in the outer blocks. That is, there are $116$ points along an equatorial line 
around each black hole and twice that number in the distant wave region. This is somehow inefficient 
since the solution is over-resolved in the angular directions compared to the radial one (especially in the 
wave region). The motivation behind this choice was 
to allow the grid points at the boundary faces of adjacent blocks to be in one-to-one
correspondence with each other. In this way the communication 
of the characteristic modes at the inter-patch boundaries does not require 
interpolation. 

In Table~\ref{tab:resolution} 
we show the total number of points
in the whole domain and per block for the simulations of this paper. We increase resolution only along
the radial direction, by the same number of points in all the
domains. In our setup all blocks
have the same number of points. Since parallelization is handled by
assigning one block per processor, this guarantees a homogeneous
load distribution. 

The number of points shown in Table~\ref{tab:resolution} is actually not large for a fully three-dimensional (i.e. 
no symmetries imposed) finite-difference simulation. For example, 
we can compare these numbers to a binary black hole evolution with around the same number of orbits  
using Cartesian grids and adaptive mesh refinement \cite{FH}. A typical state-of-the-art
simulation uses six refinement levels around each black hole with $64^3$ points on each level, and four coarse grid levels with $128^3$ points.
This amounts to a total of $2\times 6 \times 64^3 + 4 \times 128^3 \approx 226^3$ points. 
In the case of non-spinning, equal-mass black holes one can make
 use of the symmetry of the problem and reduce the total number of points to $6 \times 64^3 + 128^3 \approx 154^3$.

\begin{figure}
\includegraphics[width=0.5\textwidth]{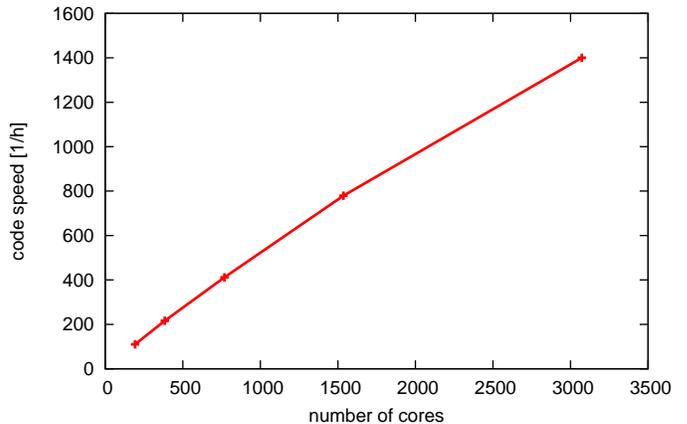}
\caption{Strong scaling test for a single black hole. The speed of the
code depends essentially linearly on the number of processors, almost
perfect scaling.}
\label{fig:strong-scaling}
\end{figure}

\begin{figure}
\includegraphics[width=0.5\textwidth]{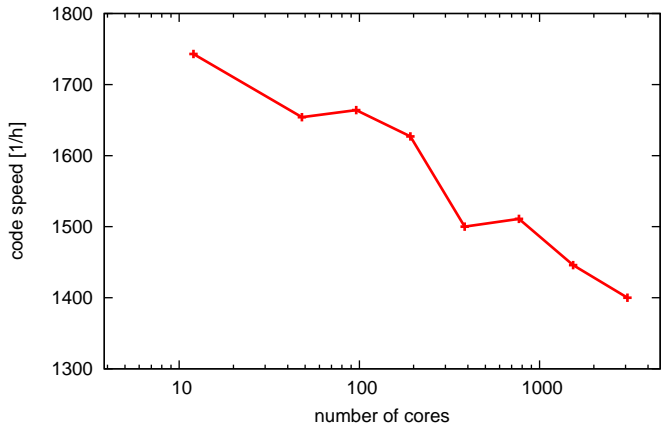}
\caption{Weak scaling test for a single black hole. There is only a 15\%
drop in speed as the number of processors is increased while keeping
the load per processor fixed.}
\label{fig:weak-scaling}
\end{figure}

We have tested the
performance of our multipatch parallelization scheme for the evolution of
a single black hole. In Fig.~\ref{fig:strong-scaling} we show
a strong scaling test for up to $3,000$ processors (cores), in which the total number of points is
kept fixed while increasing the number of processors. We see that 
the speed of the code has a linear dependence on the number of
processors. Similarly, in Fig.~\ref{fig:weak-scaling} we show
a weak scaling test, where the load per processor is kept fixed
while increasing the number of processors used. The drop in
speed in this case is about $15 \%$ over a range of 10 to
3,000 processors. We have not attempted to go beyond this number of cores. 

The phase errors in the waveforms shown in the next section are rather large compared to state-of-the-art simulations (in particular, compared 
to an AMR one such  as the one mentioned above). 
Since the code scales well and the number of points used in this paper (shown in Table~\ref{tab:resolution}) 
is reasonable for a finite-difference evolution, in principle we could improve the accuracy of the simulations shown 
in the next section while still using modest computational resources. 
What has prevented us from doing so is a purely technical obstacle. The computational infrastructure used in this paper, 
SpEC, was originally designed for pseudo-spectral evolutions, which are extremely efficient in terms of memory. For that 
reason SpEC currently stores in memory many more variables than are actually needed for evolving the system. As 
a result, in our FD simulations because of memory constraints we actually end up using a few cores per node and a rather large 
number of nodes. We plan to improve SpEC's use of memory soon to eliminate this limitation.  However, for the demonstrations
in this paper, the current resolutions are sufficient.

\begin{table}
\centering
\begin{tabular}{ccr}
\hline
$N_{r} \times N_{\rm ang} \times N_{\rm blocks} = N_{\rm total} $& speed (h$^{-1})$ & CPU (h) \\
\hline
$19 \times 29^2 \times 192 = 145^3$ & 2.83 & 67844 \\
$22 \times 29^2 \times 192 = 153^3$ & 1.86 &103226 \\
$16 \times 29^2 \times 384 = 173^3$ & 2.42 &158678 \\
\hline
\end{tabular}
\caption{Speed and CPU time for three resolutions. $N_r$ and $N_{\rm ang}$ are
the radial and angular number of points per block, respectively, as described in the text. The speed
is expressed in units of the total irreducible mass per hour.}
\label{tab:resolution}
\end{table}

\section{Results}
\begin{figure}
\includegraphics[width=0.5\textwidth]{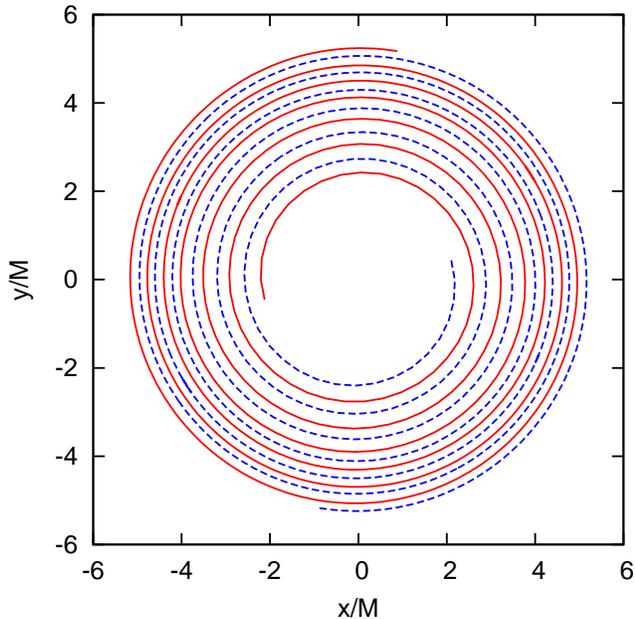}
\caption{Black hole orbits.}
\label{fig:orbits}
\end{figure}

Figure~\ref{fig:orbits} shows the location of the centroids of the apparent horizons for the highest resolution simulation. 
The black holes complete about six orbits before reaching the merger regime. 

\subsection{Convergence of the constraints}
\begin{figure}
\includegraphics[width=0.5\textwidth]{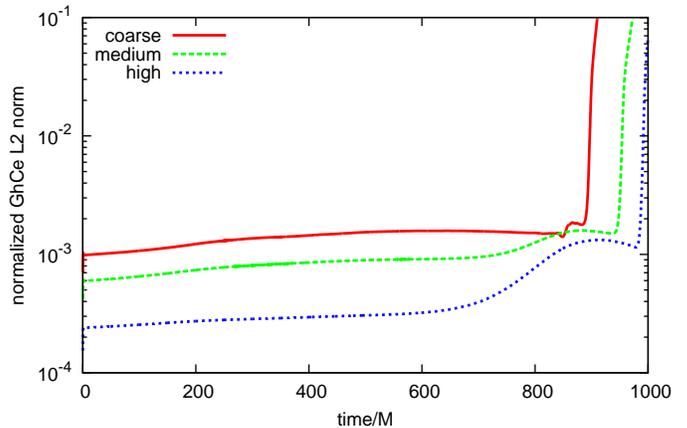}
\caption{$L^2$ norm of the normalized constraints.}
\label{fig:ghce-norm}
\end{figure}

A way of checking the consistency of the numerical solution is
monitoring the constraint violations, since they are not enforced
during the evolution. In Fig.~\ref{fig:ghce-norm} we plot the
$L^2$ norm of all the constraint fields of the first order
generalized harmonic system, normalized by the $L^2$
norm of the spatial gradients of the dynamical fields, as
defined in \cite{Lindblom:2005qh}. We show three runs with
different resolutions.

\begin{figure}
\includegraphics[width=0.5\textwidth]{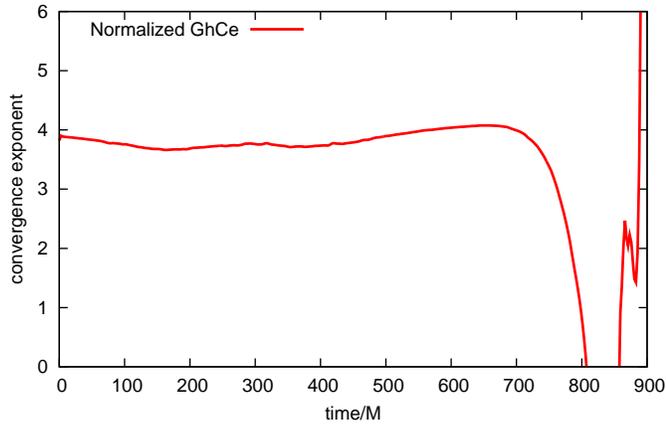}
\caption{Convergence exponent for the $L^2$ norm of the 
normalized constraints.}
\label{fig:ghce-conv}
\end{figure}

Figure~\ref{fig:ghce-conv} shows the convergence
exponent of the $L^2$ norm of the normalized constraint
violations, which is around four, as expected (cf. Sec.~\ref{sec:num}).
The convergence exponent $n$ is defined as
\beq
\frac{\beta^n-1}{\beta^n-\alpha^n}=\frac{C_1-C_3}{C_2-C_3}
\eeq
where $\alpha$ is the ratio between the medium and coarse resolution
and $\beta$, the ratio between the fine and coarse one. 
$C_1$, $C_2$, and $C_3$ represent a given quantity at coarse,
medium and fine resolutions, respectively.
 
The uniform convergence is lost around 
$t\sim 800\,M$, at which time the values for the
coarse and medium resolutions intersect, as is seen in
Fig.~\ref{fig:ghce-norm}.

We stop our simulations when the characteristic speeds at
the excision boundary change sign, which means that there is
spurious information entering the domain. That moment is characterized
by a blow-up of the constraints. This feature is due to the
inadequacy of the rather simple gauge conditions used here at times close to merger. At the time the simulations of this paper 
were performed, we used the same simple conditions used then by the \CC\ collaboration, namely, keeping the gauge source 
functions fixed in the comoving frame. Since then, better conditions have been developed, which do allow simulations to go through merger 
and ringdown \cite{Scheel:2008rj}. For the purposes of this paper, however,
following six orbits of an inspiral is sufficient.

\subsection{Waveforms}

Waveforms are computed via the Newman-Penrose curvature scalar
$\Psi_4$ as in~\cite{Pfeiffer:2007yz}. Subsequently
we decompose $\Psi_4$ in spin-weighted
spherical harmonics $_{-2}Y_{\ell m}(\theta,\phi)$. 
We focus our discussion to the $\ell=2$,
$m=2$ mode. The extraction is done at $r=50\,M$.

\begin{figure}
\includegraphics[width=0.5\textwidth]{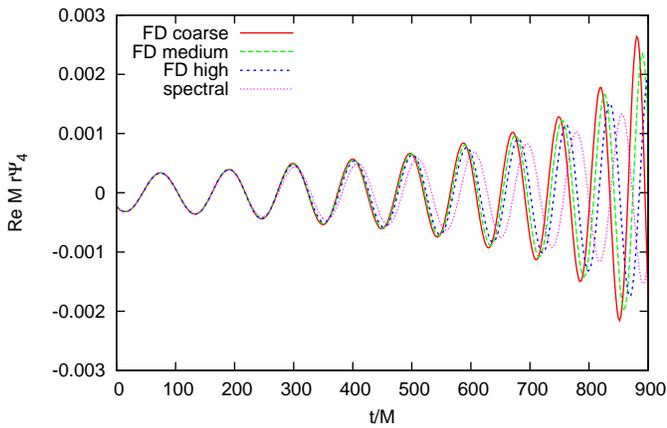}
\caption{Real part of $\Psi_4$ for the finite difference and
spectral results.}
\label{fig:psi4-re}
\end{figure}

Figure \ref{fig:psi4-re} shows the real component of $\Psi_4$ 

\begin{figure}
\includegraphics[width=0.5\textwidth]{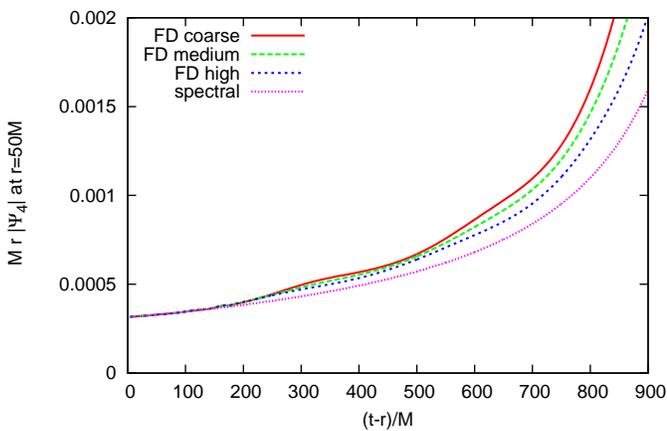}
\caption{$\Psi_4$ amplitude for the finite difference and
spectral results.}
\label{fig:psi4-amp}
\end{figure}

\begin{figure}
\includegraphics[width=0.5\textwidth]{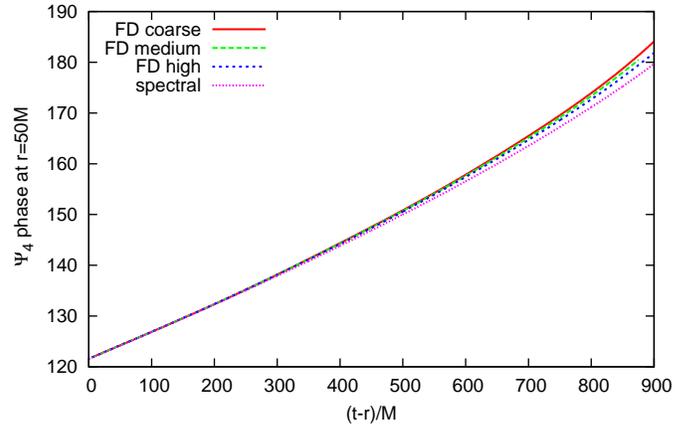}
\caption{$\Psi_4$ phase for the finite difference and
spectral results.}
\label{fig:psi4-pha}
\end{figure}

\begin{figure}
\includegraphics[width=0.5\textwidth]{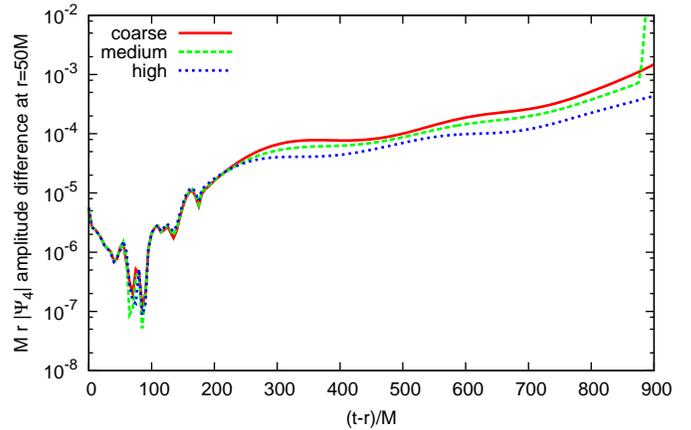}
\caption{Differences in the $\Psi_4$ amplitude between the finite 
difference and the spectral results.}
\label{fig:psi4-amp-diff}
\end{figure}

\begin{figure}
\includegraphics[width=0.5\textwidth]{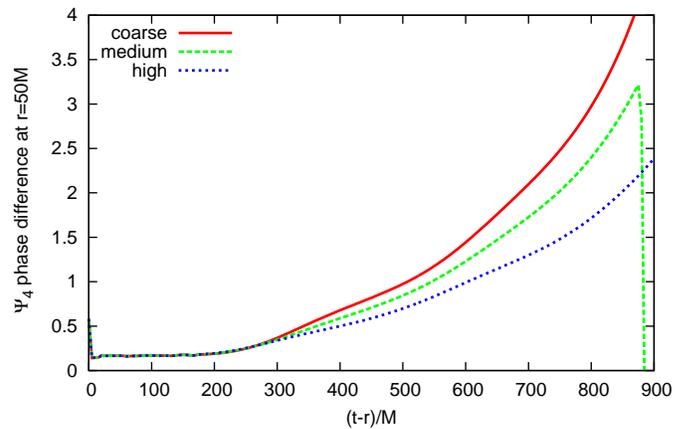}
\caption{Differences in the $\Psi_4$ phase between the finite 
difference and the spectral results.}
\label{fig:psi4-pha-diff}
\end{figure}

We see that they all agree at early times and drift apart during the
later stages of the evolution. A more meaningful comparison is 
shown in Figs.~\ref{fig:psi4-amp} and \ref{fig:psi4-pha}, where
we plot amplitude and phase of the extracted wave. 
The differences between the finite differences waveforms and the
spectral one are shown in Figs.~\ref{fig:psi4-amp-diff} and
\ref{fig:psi4-pha-diff} for the amplitude and phase, respectively.

\section{Remarks} \label{sec:remarks}

In this paper we have shown that we can evolve orbiting black holes in a stable way using a high-order multipatch approach and 
that this method scales well with the number of processors. As a result, we expect to be able to achieve 
good accuracy while still using only modest computational resources. These results also suggest that multipatch methods are an excellent alternative to
AMR, with benefits of  simplicity and $O(N)$ scaling for improving
resolution in the wave zone. Finally, multipatch methods will
allow one to use the same grid to evolve both metric and matter 
fields for a binary pair composed of a black hole and neutron star,
allowing the advantages of high-order methods without the
drawbacks of a hybrid spectral-finite difference approach.

\begin{acknowledgments}

The numerical simulations presented here were performed using the 
Spectral Einstein Code (SpEC) developed at Caltech and Cornell primarily 
by Larry Kidder, Mark Scheel and Harald Pfeiffer.

This research has been supported in part by NSF Grant No.\ PHY-0801213
to the University of Maryland, by
NSF Grants No.\ PHY-0652952, No. DMS-0553677, and No. PHY-0652929  and by
a grant from the Sherman Fairchild Foundation to Cornell, by the
TeraGrid allocation TG-MCA02N014 to Louisiana State University, and
by supercomputing resources at LONI. 

We thank Oleg Korobkin for his help on the early stages of this project.

\end{acknowledgments}

\bibliography{references}

\end{document}